\documentclass[traditabstract]{aa} 

\usepackage{graphicx}
\usepackage{amsmath}
\usepackage{bm}
\usepackage[USenglish]{isodate}
\usepackage{natbib,twoopt}
\usepackage{hyperref} 
\usepackage{color}
\bibpunct{(}{)}{;}{a}{}{,} 

\newcommand{\treifull}{3I\slash ATLAS}
\newcommand{\trei}{3I}

\newcommand{\fitepoch}{2026 February 19}
\newcommand{\epochtotal}{8339}
\newcommand{\perihelion}{2025 October 29}
\newcommand{\firstepoch}{2025 May 8}
\newcommand{\lastepoch}{2026 April 14}
\newcommand{\retrieved}{2026 June 9}

\def\aj{{AJ}} 
\def\araa{{ARA\&A}}           
               
\def\apjl{{ApJ}}              
\def\apjs{{ApJS}}           
\def\aap{{A\&A}}

\def\mnras{{MNRAS}}     
\def\nat{{Nature}}        

\def\psj{{PSJ}}

\begin{document}

\title{Systematic and Statistical Uncertainties in the Non-Gravitational Acceleration of 3I/ATLAS}

\titlerunning{Uncertainty in the Non-Gravitational Acceleration of 3I/ATLAS}

\author{
Federico Spada \inst{1}
\and
Ma\l{}gorzata Kr\'olikowska\inst{2}
\and
Luke Dones\inst{3}
}

\authorrunning{F. Spada et al.}

\institute{
Universit\`a di Catania, Dipartimento di Fisica e Astronomia, Sezione Astrofisica, Via S. Sofia 78, 95123 Catania, Italy
\email{federico.spada@gmail.com}
\and
Centrum Bada\'n Kosmicznych Polskiej Akademii Nauk (CBK PAN), Bartycka 18A, 00-716 Warszawa, Poland
\and
Department of Space Studies, Southwest Research Institute, 1301 Walnut St Suite 400, Boulder, CO 80302, USA
}

\date{}

\abstract{
We present a detailed analysis of the trajectory of interstellar comet 3I/ATLAS, with emphasis on the determination of its non-gravitational acceleration (NGA) parameters and their associated uncertainties. 
Using astrometric observations spanning both pre- and post-perihelion phases, we compute orbital solutions assuming gravity-only, symmetric, delayed-peak, and asymmetric outgassing NGA models. 
To mitigate the effects of heterogeneous astrometric quality and potential tailward biases, we adopt a tiered uncertainty model that inflates the uncertainties of lower-fidelity observations, and assess solution robustness through validation against non-fitted data.
We find clear evidence that NGA is required to reproduce the observations. 
Solutions obtained from pre- and post-perihelion arcs separately, however, yield mutually incompatible NGA parameters, suggesting either sensitivity to residual astrometric systematics or intrinsic limitations of constant parameter outgassing models.
For the full data set, the inferred acceleration is in very good agreement with the latest JPL solution, supporting its use for physical interpretation and nucleus size estimates. 
Models incorporating delayed or asymmetric activity with respect to perihelion improve the fit relative to a symmetric prescription, suggesting that part of the observed asymmetry may be physical. 
We conclude that, while the NGA of 3I is robustly determined, systematic uncertainties arising from both residual astrometric biases and simplifications in the outgassing modeling prescriptions exceed the formal fitting errors, and should be taken into account when deriving physical properties such as the nucleus size.
}


\maketitle

\section{Introduction}

\treifull{} is the third interstellar object ever detected passing through the Solar System, identified by the ATLAS survey on a strongly hyperbolic trajectory that is inconsistent with a bound Solar System origin (\citealt{Denneau_ea:2025}; \citealt{Bolin_ea:2025}, \citealt{Seligman_ea:2025}, and references therein). 
Like its predecessors (1I/‘Oumuamua and 2I/Borisov), \trei{} offers a rare opportunity to directly probe the physical and dynamical properties of small bodies formed around other stars \citep{Moro-Martin:2022, Jewitt_Seligman:2023, Jewitt:2024}.  

Observations indicate comet-like activity, suggesting that volatile-driven processes play a significant role in its evolution \citep[e.g.,][]{Cordiner_ea:2025}.
Accurately estimating the non-gravitational acceleration (NGA) acting on \trei{} is therefore essential. 
Outgassing-driven forces can measurably perturb the trajectory, biasing orbital solutions if neglected or improperly modeled. 
Reliable constraints on these accelerations are crucial not only for precise orbit determination and ephemeris prediction, but also for inferring intrinsic physical properties such as volatile composition, activity asymmetry, and thermal response. 
Moreover, robust modeling of non-gravitational effects is necessary to confidently reconstruct the object’s incoming asymptotic trajectory, which underpins any attempt to link \trei{} to a specific Galactic population or stellar birthplace \citep{Hopkins_ea:2025, Taylor_Seligman:2025, Guo_ea:2025, Albrow_ea:2025}.

Recent measurements of the ${\rm D/H}$ and $^{12}\mathrm{C}/^{13}\mathrm{C}$ isotopic ratios in \trei{} provide unprecedented insight into the pristine composition of interstellar icy planetesimals, which are found to differ markedly from known Solar System counterparts \citep{Cordiner_ea:2026, SalazarManzano_ea:2026}. 
Such extreme isotopic signatures indicate formation at very low temperatures ($< 30$ K) in a relatively metal-poor environment, early in the history of the Galaxy. These results provide independent evidence that \trei{} originates from a planetary system formed under physical and chemical conditions distinct from those of the Solar System, and belongs to a very old population with an age of order $10$--$12$~Gyr.

NGA estimates also bear directly on size determinations of the nucleus. 
To place \trei{} in context, \citet{Jewitt_Seligman:2023} estimated that 2I had a diameter of $\sim 0.4$--$1$ km, assuming a geometric albedo of $p = 0.1$. Its dust mass-loss rate at $2$ au was reported to be $\sim 70$ kg s$^{-1}$, although this estimate is likely uncertain at the order-of-magnitude level \citep{Kim_ea:2020}. 
The JPL orbital solution for 2I yields a magnitude of the NGA normalized at $1$ au of approximately $5.3 \cdot 10^{-8}$ au d$^{-2}$, assuming a $1/r^2$ radial decay of the acceleration, appropriate for outgassing of substances more volatile than water.
By contrast, 1I exhibited an NGA approximately five times larger, detectable at the $\sim 30\sigma$ level \citep{Micheli_ea:2018}, but in spite of extensive searches, no cometary activity was detected \citep[e.g.,][]{Jewitt_ea:2017, Meech_ea:2017, Ye_ea:2017, Fitzsimmons_ea:2018}.

For \trei{}, reported dust production rates are of order $\sim 180$ kg s$^{-1}$ at $2$ au \citep{Jewitt_Luu:2025}.
The JPL orbital solution \#54 for \trei{} yields a magnitude of the NGA at $1$ au of approximately $5.5 \times 10^{-8}$ au d$^{-2}$, under the assumption that the activity is driven by something more volatile than water, such as CO$_2$. 
Assuming water-driven activity, on the other hand, \citet{Ahuja_Ganesh:2025} obtained a NGA of $\approx 3 \cdot 10^{-6}$ au d$^{-2}$, which seems implausibly large.
The similarity in dust production rates and NGA of 2I and \trei{} therefore suggest that these objects are likely comparable in size, of the order of $1$ km in diameter \citep{Forbes_Butler:2026}, although a substantially larger early estimate was proposed by \citet{Cloete_ea:2025} based on now-outdated upper limits on the NGA. 

Independent constraints from surface-brightness limits and sublimation modeling further support a kilometer-scale nucleus for 3I, with the preferred size depending on the dominant volatile and dust grain properties \citep{Jewitt_ea:2025, Cordiner_ea:2025}.
Most recently, several determinations of the size of \trei{}, based on different methods, were summarized by \citet{Hui_ea:2026}, who recommend a diameter of $\approx 3$ km on the basis of both nucleus extraction techniques and NGA estimates.

The aim of this work is to assess the impact of astrometric uncertainty, modeling, and data selection on the determination of the NGA parameters for \trei{}, using publicly available astrometric data.
In particular, we demonstrate the sensitivity of the inferred NGA parameters to data selection and uncertainty assignment, and we quantify the resulting variations.
In addition to models based on the standard $g(r)$ activity function \citep{Marsden_ea:1973, Yeomans_Chodas:1989}, we also introduce a more general formulation in which the dependence of the NGA on the heliocentric distance is described by different power-law exponents before and after perihelion, thereby relaxing the usual symmetry assumption.

This paper is organized as follows. The observational data used to constrain the trajectory of \trei{} are described in Section~\ref{data}. 
Section~\ref{methods} outlines the orbit determination approach and the assessment of parameter uncertainties. 
Results are presented in Section~\ref{results} and discussed in Section~\ref{discussion}, with conclusions summarized in Section~\ref{conclusions}.

\section{Observational data}
\label{data}

\subsection{The astrometric dataset}

The astrometric dataset used in this work to constrain the trajectory of \trei{} consists of optical measurements of its sky position, expressed as right ascension $\alpha$ and declination $\delta$ in the J2000 equatorial reference frame, obtained over \epochtotal{} epochs between \firstepoch{} and \lastepoch{}.
The full dataset is publicly available and was retrieved from the Minor Planet Center (MPC) web interface\footnote{\url{https://data.minorplanetcenter.net/explorer/?tab=Designated&search=3I}}
 on \retrieved{}.
Observations included in the MPC dataset were obtained from $343$ different stations.

The astrometric measurements of \trei{} were acquired by a combination of ground- and space-based facilities.
As of \retrieved, the publicly available space-based measurements include pre-discovery archival frames from the Transiting Exoplanet Survey Satellite (TESS) obtained in May 2025 \citep{Martinez-Palomera_ea:2025}; Hubble Space Telescope (HST) observations on 2025 July 21 \citep{Jewitt_ea:2025}; and target-of-opportunity imaging by the interplanetary spacecraft Psyche and Trace Gas Orbiter (TGO).

\begin{figure}
\begin{center}
\includegraphics[width=0.5\textwidth]{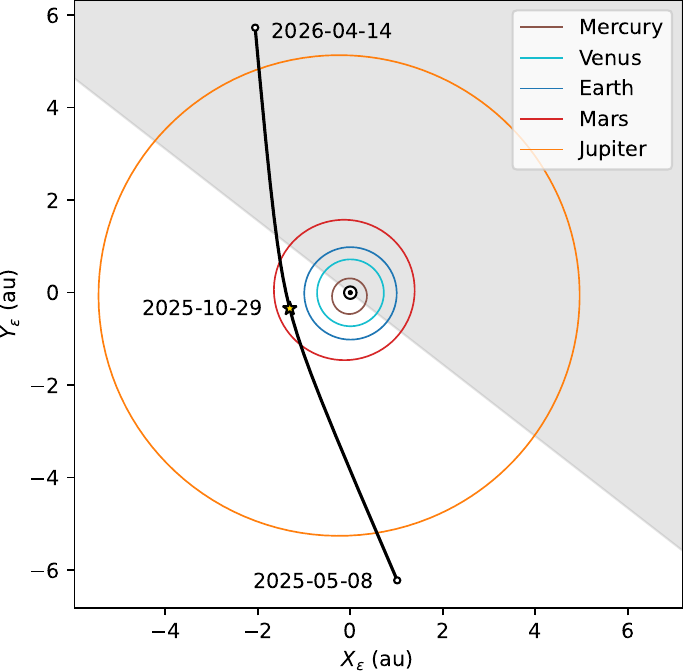}
\caption{Orbit of \trei{} in the heliocentric ecliptic reference frame. 
Perihelion is marked by a yellow star, and the endpoints of the observational arc used for the orbit determination in this study are indicated by open circles. 
Gray shading denotes the region where the trajectory lies above the ecliptic plane. 
The Sun and the orbits of the major planets through Jupiter are shown for reference.}
\label{fig:orbit}
\end{center}
\end{figure}

Figure~\ref{fig:orbit} shows the orbit of \trei{} in the heliocentric ecliptic frame, highlighting the portion of the trajectory corresponding to the observational arc used for orbit determination in this work (see Section~\ref{orbfits}).
\trei{} reached perihelion on 2025 October 29 at a distance of $q \approx 1.356$ au.
A passage near Jupiter, just outside its Hill radius, occurred on 2026 March 16.

With an eccentricity of $e \approx 6.1$, the trajectory is strongly hyperbolic.
Notably, \trei{} had the largest hyperbolic excess velocity ($v_\infty \approx 58$ km s$^{-1}$) among the interstellar objects detected in the Solar System to date.

\begin{table*}
\centering
\caption{Observational uncertainty model used for orbit determination in this work; the uncertainty inflation factor $s$ applied to Tier~3 observations is discussed in the main text.}
\begin{tabular}{cclccc}
\hline
\hline
Tier & Identifier & Name / Notes & Uncertanity & Rejection & Number of \\
 & & & inflation factor  & allowed? & observations \\
\hline
 1 &  250 & Hubble Space Telescope &    1   &  No &    4 \\
 1 &  309 &        Cerro Paranal &    1   &  No &   43 \\
 1 &  705 &         Apache Point &    1   &  No &   11 \\
 1 &  I33 &   SOAR, Cerro Pachon &    1   &  No &    6 \\
 1 &  T14 & Canada-France-Hawaii Telescope, Maunakea &    1   &  No &    9 \\
 1 &  T15 & Gemini North Observatory, Maunakea &    1   &  No &   12 \\
 1 &  339 &    Trace Gas Orbiter &    1   &  No &    4 \\
 1 &  336 &                 Lucy &    1   &  No &    3 \\
 1 &  338 &               Psyche &    1   &  No &    2 \\
 \hline
 2 &  703 &  Catalina Sky Survey &   1   &  Yes &   56 \\
 2 &  E10 & Siding Spring-Faulkes Telescope South &   1   &  Yes &   50 \\
 2 &  F51 & Pan-STARRS 1, Haleakala &   1   &  Yes &    3 \\
 2 &  F52 & Pan-STARRS 2, Haleakala &   1   &  Yes &    4 \\
 2 &  F65 & Haleakala-Faulkes Telescope North &   1   &  Yes &   83 \\
 2 &  G96 &    Mt. Lemmon Survey &   1   &  Yes &   17 \\
 2 &  I41 & Palomar Mountain--ZTF &   1   &  Yes &   77 \\
 2 &  J13 & La Palma-Liverpool Telescope &   1   &  Yes &   59 \\
 2 &  T05 & ATLAS-HKO, Haleakala &   1   &  Yes &  207 \\
 2 &  T08 & ATLAS-MLO, Mauna Loa &   1   &  Yes &   46 \\
 2 &  W84 &   Cerro Tololo-DECam &   1   &  Yes &  137 \\
 2 & --- & Zero-aperture extrapolated, seeing < 1.0 arcsec & 1 & Yes &   17 \\
 \hline
 3 & --- & All remaining observations & $s$ & Yes & 7489 \\
\hline
\end{tabular}
\label{tab:data_tiers}
\end{table*}

\subsection{Treatment of observational uncertainties}
 \label{downweighting}
 
Since its discovery, \trei{} has exhibited a characteristically cometary appearance.
Astrometry of active comets is notoriously difficult to perform accurately because the asymmetric light distribution around the nucleus can introduce a systematic offset in the measured photocenter position (tailward bias; see, e.g., \citealt{Farnocchia_ea:2016}).

For \trei{}, this issue is compounded by the extreme heterogeneity of the dataset, which was acquired by a large variety of facilities operated by observers with widely differing levels of experience.
The brightness of \trei{} during the months surrounding perihelion, together with its obvious significance as only the third recognized interstellar object to date, led to an unusually large volume of astrometry from small telescopes operated by amateur observers.
At the same time, non-gravitational accelerations, which are the main focus of this work, are comparatively subtle effects whose determination can be strongly biased by tailward systematics.
For these reasons, the assignment of realistic uncertainties to the astrometry of \trei{} is an essential pre-condition for reliable orbit determination.

The most rigorous treatment of tailward bias would require a case-by-case assessment of each observing facility and, in some instances, an independent remeasurement of astrometric positions from the original raw images \citep[e.g.,][]{Farnocchia_ea:2016}.
Given the breadth of contributing observers, such an effort would be highly time-consuming and lies beyond the scope of the present work.
We therefore adopted a heuristic classification scheme in which the full dataset is divided into subsets with different expected levels of reliability.

The guiding principle is to minimize explicit outlier rejection while increasing, from the outset, the assigned uncertainties of observations expected to be most affected by systematic bias. 
In this way, the highest quality measurements, although fewer in number, are less likely to be discarded, whereas lower quality data may remain in the fit with appropriately reduced statistical influence. 
Our procedure is as follows.

As a starting point, all observations are assigned the observer-reported random uncertainties listed in the ADES-formatted records retrieved from the MPC web interface, namely \texttt{rmsRA} for $\sigma_{\alpha^*} = \sigma_{\alpha}\cos\delta$, \texttt{rmsDec} for $\sigma_{\delta}$, and \texttt{rmsCorr} for the correlation coefficient $\rho$, when available.
When these quantities are not provided, we adopt the default values $\sigma_{\alpha^*} = \sigma_{\delta} = 1.0$ arcsec and $\rho = 0$.

Besides tailward-bias systematics, observer-reported astrometric uncertainties are often underestimated and therefore cannot always be regarded as reliable. 
We therefore introduce a tiered classification scheme that reflects the expected reliability of different observations and governs their treatment during the fitting procedure, including rejection (see Section~\ref{orbdet}) and uncertainty inflation (cf. Table~\ref{tab:data_tiers}):

\begin{itemize}
\item Tier 1 comprises observations from HST, the Very Large Telescope, and other ground-based facilities of comparable class, together with the measurements obtained from the deep-space probes Lucy, Psyche, and TGO (94 observations in total). 
These observations are never considered for rejection during the orbit determination procedure, and their reported uncertainties are used without modification.
\item Tier 2 includes observations from professional surveys and large telescopes, as well as measurements acquired with the zero-aperture extrapolation technique \citep{Tholen_Chesley:2004}, but only if their reported seeing is better than $1$ arcsec (756 observations in total).
Tier 2 observations are considered for rejection during the fitting procedure, 
but their observer-reported uncertainties are not inflated.
\item Tier 3 consists of all remaining observations in the dataset ($7489$ observations). 
 Their observer-reported uncertainties are multiplied by a factor $s > 1$. Increasing $s$ progressively reduces the influence of Tier~3 observations on the fit, and in the limit of large $s$ the solution approaches that obtained using only Tier~1 and Tier~2 observations. 
We explored the sensitivity of the orbital solutions, and in particular of the fitted NGA parameters, over the range ($1 \le s \le 25$). 
Based on this analysis, we adopt $s=15$, for which the full-arc solution is already close to its asymptotic limit (see Section~\ref{deweight_sensitivity}).
\end{itemize}

A further source of underestimated formal uncertainty arises from over-observation, i.e., large numbers of measurements obtained during the same night from the same station. 
To mitigate this effect, when $N > 4$ observations are available from a given station on a single night, their assigned uncertainties are further multiplied by the factor $\sqrt{N/4}$ \citep[e.g.,][]{Veres_ea:2017}.

The normalized residuals used in the orbit fitting procedure and shown in the diagnostic plots are defined, for both right ascension and declination, as
\begin{equation}
r_{\mathrm{norm}}=\frac{r_{\mathrm{abs}}}{\sigma_{\mathrm{eff}}},
\end{equation}
where $r_{\mathrm{abs}}$ is the observed-minus-computed (O$-$C) astrometric residual in arcseconds, and $\sigma_{\mathrm{eff}}$ is the effective uncertainty, i.e., after applying the uncertainty inflation factors described above. 
Consequently, the normalized residuals are dimensionless, whereas both the residuals and effective uncertainties have units of arcseconds.

For Tier~1 and Tier~2 observations, $\sigma_{\mathrm{eff}}$ is equal to the observer-reported uncertainty, apart from the additional inflation applied to account for over-observation, if applicable. 
For Tier~3 observations, $\sigma_{\mathrm{eff}}$ includes both the adopted factor $s$ and, when appropriate, the over-observation correction.

It should be noted that tailward bias is a systematic effect and therefore cannot be eliminated through uncertainty scaling, which addresses only random error contributions. 
Nevertheless, the adopted procedure reduces its impact and partially mitigates the resulting systematic errors.

\section{Methods}
\label{methods}

\subsection{Dynamical model}
\label{dynamics}

Our dynamical model of the trajectory of \trei{} is based on the numerical integration of the equations of motion expressed in Cartesian coordinates within a heliocentric equatorial reference frame referred to the J2000 equator and equinox.
The total acceleration acting on the body is expressed as the sum of a gravitational and a non-gravitational component.

The gravitational part of the acceleration includes the central term from the Sun and perturbations from the major planets, Pluto, and the Moon, as well as the $16$ most massive asteroids, listed in Table 2 of \citet{Farnocchia_ea:2015}.
The general relativity correction term for solar gravity is also included, according to the post-Newtonian approximation \citep{Damour_Deruelle:1985}:
\begin{equation}
{\bm a}_{\rm REL} = \frac{\mu_\odot}{c^2 r^3}\left[ \left(\frac{4 \mu_\odot}{r} - v^2 \right) {\bf r} + 4 ( {\bf r} \cdot {\bf v}) {\bf v} \right],
\end{equation}
where $\mu_\odot$ is the gravitational parameter of the Sun.

The non-gravitational component of the acceleration is modeled following the \citet{Marsden_ea:1973} formalism:
\begin{equation}
\label{aNG}
{\bf a}_{\rm{NG}} = g(r) \, \left[ A_1 {\bf e}_R + A_2 {\bf e}_T + A_3 {\bf e}_N \right].
\end{equation}
In the equation above, $A_1, A_2, A_3$ are the NGA parameters along the radial, transverse, and normal directions of the standard RTN (radial--transverse--normal) frame. 
The unit vector ${\bf e}_R$ is along the instantaneous radial direction; ${\bf e}_N$ is normal to the instantaneous orbital plane, and ${\bf e}_T$ lies in that plane and is orthogonal to ${\bf e}_R$, forming a right-handed triad.

The function $g(r)$ describes the dependence on the heliocentric distance of the outgassing-induced acceleration. 
We adopt the parametrization introduced by \citet{Marsden_ea:1973} as our reference:
\begin{equation}
\label{g_func}
g(r) =
\alpha
\left( \frac{r}{r_0} \right)^{-m}
\left[
1 + \left( \frac{r}{r_0} \right)^n
\right]^{-k},
\end{equation}
where $\alpha$, $r_0$, $m$, $n$, and $k$ are constant parameters.
Throughout this work we adopt $\alpha = 1$, $r_0 = 1$ au, $m = 2$, and $n = k = 0$, corresponding to $g(r)=(1\, {\rm au} /r)^2$.
This choice is motivated by the hypothesis that the activity of \trei{} is driven by highly volatile species such as CO or CO$_2$ \citep{Cordiner_ea:2025}.

In the standard formulation, the NGA depends only on the instantaneous heliocentric distance, i.e., $g[r(t)]$.
The resulting activity profile therefore reaches its maximum at perihelion, and is symmetric about it.

A commonly adopted extension introduces a temporal asymmetry by evaluating the activity law at a delayed time:
\begin{equation}
\label{g_delay}
g[r(t-DT)],
\end{equation}
where the parameter $DT$ shifts the epoch of peak activity before ($DT < 0$) or after ($DT > 0$) perihelion \citep{Yeomans_Chodas:1989}.
This formulation preserves the same radial scaling before and after perihelion, while allowing the peak outgassing to occur at a different epoch.

Motivated by observational evidence that the activity of \trei{} evolves differently before and after perihelion \citep{Hui_ea:2026, Tan_ea:2026}, we also consider an alternative parameterization in which the radial decay slope itself changes across perihelion:
\begin{equation}
\label{g_nga}
g_1(r) =
\alpha \left( \frac{r}{r_0} \right)^{-m}
 \left[(1-S) \left( \frac{r}{r_0} \right)^{+\delta_m} + S \left( \frac{r}{r_0} \right)^{-\delta_m} \right].
\end{equation}
The transition function
\begin{equation}
\label{g_switch}
S(r) =\frac{1}{2} \left[1 + \tanh \left( w \frac{{\bf r}\cdot{\bf v}}{rv} \right) \right]
\end{equation}
smoothly switches between the pre- and post-perihelion regimes, with the parameter $w = 0.05$ controlling the sharpness of the transition.
Far from perihelion, the effective radial dependence thus becomes:
\begin{equation}
g_1(r) \propto
\left\{
\begin{array}{ll}
r^{-m+\delta_m}, & \text{pre-perihelion}, \\[6pt]
r^{-m-\delta_m}, & \text{post-perihelion}.
\end{array}
\right.
\end{equation}
Unlike the delayed-time prescription, this formulation introduces an intrinsic asymmetry in the heliocentric-distance scaling itself.

In this work we consider the following classes of dynamical models:

\begin{enumerate}
\item
{gr0:} purely gravitational dynamics.
\item
{ng1:} symmetric non-gravitational model, based on Eq.~\eqref{g_func}.
\item
{ng2:} time-offset asymmetric model, based on Eq.~\eqref{g_delay}.
\item
{ng3:} slope-asymmetric model, based on Eq.~\eqref{g_nga}.
\end{enumerate}

\subsection{Numerical propagation of the trajectory}

The \trei{} orbital solutions presented in this work were obtained through numerical integration of the equations of motion together with the associated variational equations.
Two independent propagators are available within our computational framework.

The first is based on the \texttt{ASSIST} package \citep{Holman_ea:2023}, which provides an efficient and well-tested environment for high-precision, ephemeris-quality Solar System integrations.
\texttt{ASSIST} is an extension of the \texttt{REBOUND} framework \citep{Rein_Liu:2012} and makes use of its IAS15 integrator \citep{Rein_Spiegel:2015}.
This propagator was adopted for the orbit fits implementing the gr0 and ng1 dynamical models.

The second propagator is based on \texttt{solve\_ivp}, part of the \texttt{SciPy.integrate} library.
In this case, the equations of motions and the variational equations were integrated using a Python version of the variable order Adams-Bashforth-Moulton predictor-corrector method \citep{Shampine_Gordon:1975, Watts:1983, Watts_Shampine:1986} available in the \texttt{extensisq} package\footnote{\url{https://github.com/WRKampi/extensisq}}.
The positions of Solar System perturbers were obtained from the DE440 JPL Planetary and Lunar Ephemerides \citep{Park_ea:2021}, accessed through NAIF/SPICE routines \citep{Acton_ea:1996,Acton_ea:2018} via the Python wrapper SpiceyPy \citep{Annex_ea:2020}.
This implementation allows fully customizable user-defined force models and was therefore used for the ng2- and ng3-type NGAs described above.

The availability of two independent propagators enables a direct cross-check of our numerical results. 
Consistency tests were performed for the gr0 and ng1 dynamical models, with the \texttt{ASSIST} propagator run with the \texttt{PLANETS}, \texttt{ASTEROIDS}, and \texttt{GR\_EIH} options enable. 
The two implementations yield mutually consistent orbital solutions when equivalent force models are adopted. 
In particular, the NGA parameters for the ng1 solution agree to within one part in $10^6$ between the \texttt{ASSIST} and the \texttt{solve\_ivp}-based propagators.

\subsection{Orbit Determination}
\label{orbdet}

We determine the orbit of \trei{} from the available astrometric data using a standard batch least-squares orbit determination procedure \citep[e.g.,][]{Milani_Gronchi:2010, Farnocchia_ea:2015}. The procedure estimates a parameter vector ${\bf x}$ by minimizing the cost function
\begin{equation}
Q({\bf x}) = {\bm \nu}^{T}({\bf x}) W {\bm \nu}({\bf x}),
\end{equation}
where ${\bm \nu}({\bf x})$ is the vector of astrometric (O$-$C) residuals corresponding to the current estimate of ${\bf x}$, and $W$ is the weight matrix, calculated as the inverse of the observational covariance matrix. 
When correlation information is available, $W$ assumes a block-diagonal form, where the block associated with observation $i$ is given by \citep[e.g.,][]{Vavilov_ea:2026}:
\begin{equation}
W_i = 
\frac{1}{(1-\rho^2)}
\left( 
\begin{array}{cc}
\sigma_{\alpha^*, i}^{-2} & -\rho_i (\sigma_{\alpha^*, i} \sigma_{\delta,i})^{-1} \\
-\rho_i (\sigma_{\alpha^*, i} \sigma_{\delta,i})^{-1} & \sigma_{\delta, i}^{-2} \\ 
\end{array}
\right)
\end{equation}
where $\sigma_{\alpha^*,i}$ and $\sigma_{\delta,i}$ are the uncertainties in right ascension and declination (assigned according to the three-tier scheme detailed in Section~\ref{downweighting}), and $\rho_i$ is their correlation coefficient.
When correlations among observations are neglected, $W$ simplifies to diagonal form, with elements given by the inverse squares of the assigned astrometric uncertainties.

The parameter vector ${\bf x}$ always includes the Cartesian heliocentric state vector $(x, y, z, v_x, v_y, v_z)$ at the reference epoch in the equatorial J2000 frame. 
In the non-gravitational model ng1, ${\bf x}$ additionally includes the three NGA parameters $A_1$, $A_2$, and $A_3$ defined in equation~\eqref{aNG}. 
In the purely gravitational model gr0, the fit is limited to the state vector components, while in model ng2 the  parameter $DT$ is included as an additional degree of freedom.
Similarly, the additional degree of freedom $\delta_m$ is included in the ng3 orbital fits (see Section~\ref{asymm}).

After linearization of the residuals with respect to the parameters, the design matrix
$B = \partial {\bm \nu} / \partial {\bf x}$ is introduced, together with the normal matrix
$C = B^{T} W B$.
The parameter vector is then iteratively updated according to
\begin{equation}
{\bf x}^{(k+1)} - {\bf x}^{(k)} = - C^{-1} B^{T} W {\bm \nu},
\end{equation}
where the covariance matrix of the estimated parameters is given by $\Gamma = C^{-1}$.
Upon convergence, ${\bf x}$ and $\Gamma$ represent the linearly estimated solution and its associated covariance matrix.

The fit was carried out using \texttt{orbit\_finder}, a Python-based tool designed for orbit determination of small Solar System bodies, publicly available on GitHub \citep{orbitfinder}.
The \texttt{orbit\_finder} pipeline implements the standard weighted least-squares orbit fit procedure described above \citep[for more details, see][]{Milani_Gronchi:2010}, with catalog bias correction according to the prescription of \citet{Eggl_ea:2020}, and the iterative outlier rejection described in \citet{Carpino_ea:2003}.

\section{Results}
\label{results}

\subsection{Orbit fits with standard dynamical models}
\label{orbfits}

The full astrometric data set used to constrain the orbit of \trei{} spans the interval from \firstepoch{} to \lastepoch{}, for a total of \epochtotal{} reported observations. 
Solar conjunction occurred during October 2025, resulting in a gap in ground-based observations of approximately $20$ days, just before the passage through perihelion on 2025 October 29.

All orbital solutions discussed below are referenced to the epoch \fitepoch{}. 
Observations acquired between \fitepoch{} and \lastepoch{} are excluded from the fits and reserved as an independent validation dataset.
The fitted and validation samples thus comprise $7950$ and $389$ observations, respectively.
Observational uncertainties were assigned as described in Section \ref{downweighting} and corrected for catalog bias according to the \citet{Eggl_ea:2020} prescription.
The automated outlier rejection procedure described by \citet{Carpino_ea:2003}, implemented in the \texttt{orbit\_finder} pipeline, was applied iteratively during the fitting process. 

\begin{figure}[h!]
\begin{center}
\includegraphics[width=0.49\textwidth]{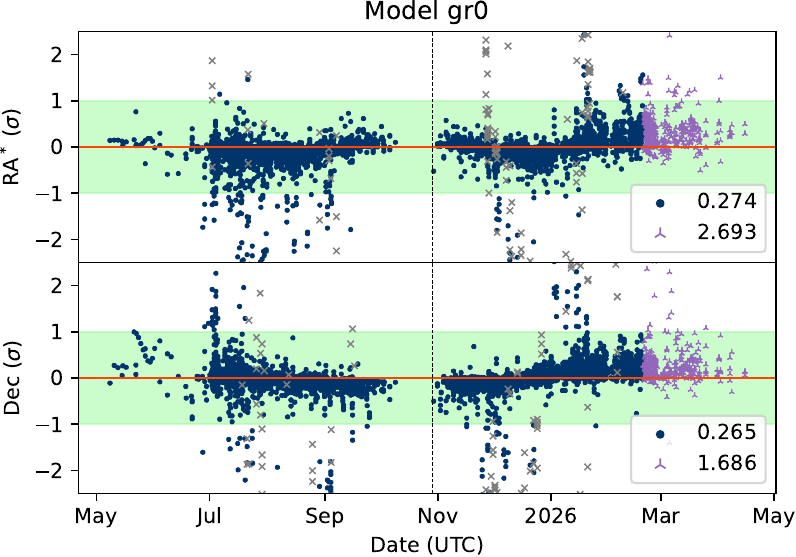}
\par\smallskip
\vspace{4mm}
\includegraphics[width=0.49\textwidth]{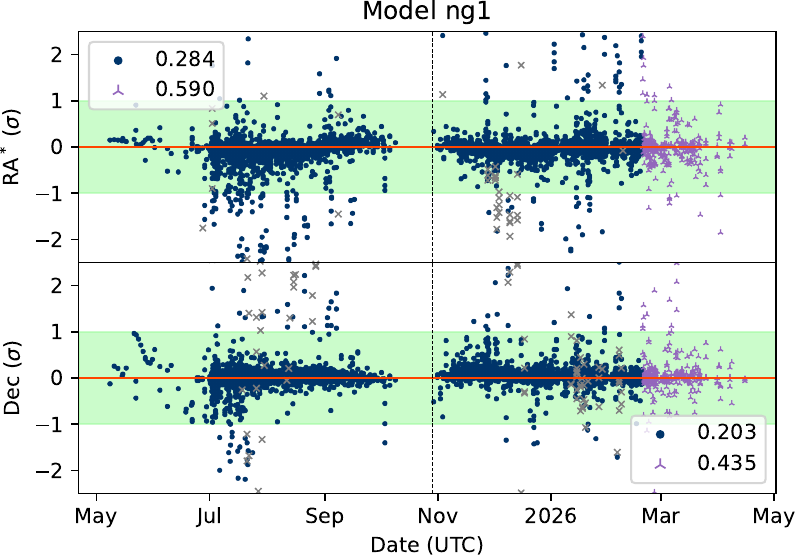}
\par\smallskip
\vspace{4mm}
\includegraphics[width=0.49\textwidth]{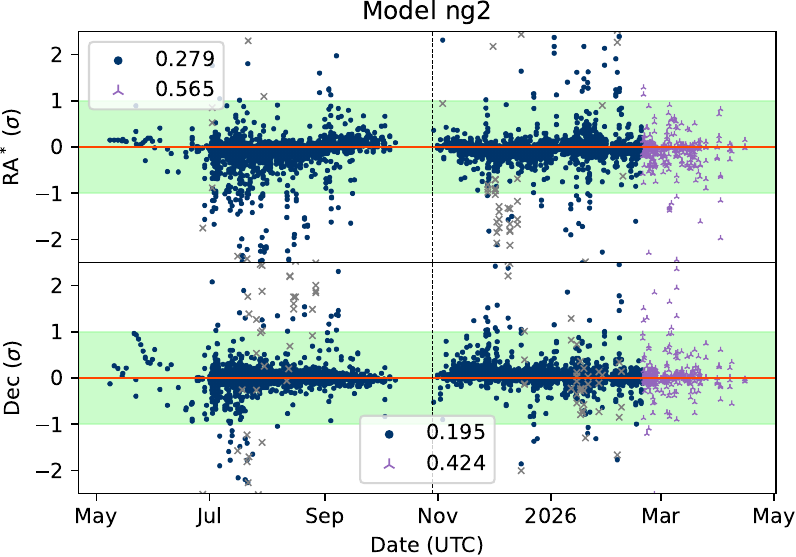}
\caption{
Normalized post-fit residuals for the orbit solutions described in the text. 
Blue dots, gray crosses, and purple upward-pointing ticks indicate fitted observations, rejected outliers, and validation data, respectively. 
RMS residuals are reported for fitted and validation samples. 
The $\pm1\sigma$ region is shown in light green. Black dashed vertical lines mark perihelion passage (2025 October 29).
}
\label{fig:residuals}
\end{center}
\end{figure}

In this section we discuss five different orbital solutions: a purely gravitational model (gr0); a symmetric non-gravitational model fitted to the full dataset (ng1); the same symmetric model fitted separately to pre-perihelion observations (ng1-pre), and to post-perihelion observations (ng1-pst); and an asymmetric non-gravitational model, in which the outgassing profile peaks at a time offset from perihelion (ng2).
The naming convention of these solutions is summarized in Table~\ref{tab:names}.
Goodness-of-fit statistics and, where applicable, NGA parameters for these solutions are listed in Table~\ref{tab:orbfits}. 
Solutions gr0, ng1, and ng2 are further compared in terms of their  post-fit residuals in Figure~\ref{fig:residuals}, while Table~\ref{tab:elts} reports their osculating heliocentric ecliptic elements and NGA parameters.
The vertical axis limits of the panels in Figure~\ref{fig:residuals} were chosen to emphasize the detailed structure and trends in the residual distributions, at the expense of truncating the most extreme outliers.
To provide a more complete view of the residuals, a version of Figure~\ref{fig:residuals} with expanded vertical limits, but otherwise identical, is included in Appendix~\ref{addfig}.

\begin{table}
\caption{Summary of the orbital solutions discussed in the text.}
\begin{center}
\begin{tabular}{cccc}
\hline
\hline
Name & $g(r)$ & NGA parameters &  Data arc \\
\hline
gr0  & ---                          & ---  & full$^*$   \\ 
ng1 & eq. \eqref{g_func}   & $A_1$, $A_2$, $A_3$   & full$^*$   \\ 
ng1-pre & eq. \eqref{g_func} & $A_1$, $A_2$, $A_3$   & pre-perihelion$^\dagger$   \\ 
ng1-pst & eq. \eqref{g_func} & $A_1$, $A_2$, $A_3$   & post-perihelion$^\ddagger$    \\ 
ng2 & eq. \eqref{g_delay} & $A_1$, $A_2$, $A_3$, $DT$   & full$^*$    \\ 
ng3 & eq. \eqref{g_nga}    & $A_1$, $A_2$, $A_3$, $\delta_m$   & full$^*$    \\ 
\hline
\end{tabular}
\end{center}
\small
$^*$: \firstepoch{} -- \fitepoch{} \\
$^\dagger$: \firstepoch{} -- \perihelion{} \\
$^\ddagger$: \perihelion{} -- \fitepoch{}
\label{tab:names}
\end{table}

\begin{table*}
\caption{Parameters of the orbital solutions discussed in Section~\ref{orbfits}.}
\begin{center}
\small
\begin{tabular}{lcccccccc}
\hline
\hline
Name &  RMS res. & $\chi^2_\nu$  & \# Obs. & $A_1$ & $A_2$ & $A_3$ & $A$ & $DT$ \\
           & (arc sec) & & included & \multicolumn{4}{c}{($10^{-8}$ au d$^{-2}$)} & (d) \\
\hline
gr0 & $ 0.850$ & $ 0.073$ & $7669$  & ---  & ---  & ---  & ---  & ---  \\
ng1 & $ 0.769$ & $ 0.061$ & $7790$  & $ 4.941 \pm 0.073$ & $ 1.801 \pm 0.089$ & $-0.762 \pm 0.010$ & $ 5.314 \pm 0.075$ & --- \\ 
ng1-pre & $ 0.771$ & $ 0.067$ & $4236$  & $ 9.305 \pm 1.186$ & $ 2.826 \pm 2.395$ & $-3.099 \pm 0.315$ & $10.206 \pm 1.272$ & --- \\ 
ng1-pst & $ 0.766$ & $ 0.052$ & $3569$  & $ 8.753 \pm 1.018$ & $-5.297 \pm 5.030$ & $-1.571 \pm 0.245$ & $10.352 \pm 2.715$ & --- \\
ng2 & $ 0.771$ & $ 0.058$ & $7794$  & $ 5.381 \pm 0.085$ & $ 1.149 \pm 0.139$ & $-0.760 \pm 0.010$ & $ 5.554 \pm 0.087$ & $   7.49 \pm  0.92$ \\ 
\hline
\end{tabular}
\end{center}
\label{tab:orbfits}
\end{table*}

Before discussing in detail the fits of Figure~\ref{fig:residuals}, we note from the outset the presence of several groups of strongly deviant observations. 
These outliers often appear independently of the adopted dynamical model, exhibiting similar patterns across all three panels of the Figure (as well as in the split-arc solutions, not shown here), and are therefore likely attributable to observational errors.
Tests in which these groups of observations are excluded and the fits repeated show no significant change in the fits, aside from the expected reduction in RMS residuals.
All these observations belong to Tier 3 in our tiered uncertainty model (see Section~\ref{downweighting}), so their influence on the fit is effectively mitigated by the inflated uncertainty. 
For this reason, they are retained in the final solutions.

It should also be noted that normalization of the residuals by their effective uncertainties can lead to a non-intuitive visual impression of the outlier distribution in diagnostic plots.
In particular, most of the rejected outliers shown in Figure~\ref{fig:residuals} belong to Tier~2 observations. 
These are excluded despite having only moderate absolute residuals (of order $\sim$1 arcsec), because their assigned uncertainties are small. 
Conversely, the majority of Tier~3 observations are retained in the fit even when they exhibit comparatively large absolute residuals, since their uncertainties are correspondingly inflated according to our uncertainty handling scheme.

The post-fit residuals obtained for the gravity-only model gr0 are shown in the top panel of Figure~\ref{fig:residuals}. 
Clear long-term trends are present in the right ascension and declination residuals, both before and after perihelion. 
Similar trends persist into the validation subsample.
These features are evidence of a significant non-gravitational component of the acceleration affecting the motion of \trei{} over the observed time span.

The residuals for model ng1 fitted to the complete dataset are shown in the middle panel of Figure~\ref{fig:residuals}. 
Comparison of the top and middle panel of the Figure shows that inclusion of the NGA removes most of the long-term trends as a function of time visible in the gr0 solution. 
Modest systematic trends remain in the vicinity of perihelion, especially in the right ascension residuals.

The remaining systematics of solution ng1 near perihelion likely arises from residual tailward biases. 
This is plausible given that \trei{} was both more active in the vicinity of perihelion and observed at lower solar elongation, conditions that may enhance systematic errors in astrometric measurements.
An additional contribution may stem from the simplifying assumptions of the adopted NGA model relative to the more complex physics of the outgassing process, particularly its directional geometry \citep[e.g.,][]{Yang_ea:2026}.

Compared with the gravity-only solution, the RMS of the residuals for the ng1 solution is lower, and the NGA parameters are tightly constrained (within a few percent). 
Overall, the comparison between models gr0 and ng1 provides strong dynamical evidence that a purely gravitational description of the motion is incomplete.

Non-symmetric outgassing with respect to perihelion may affect the determination of the NGA parameters and contribute to their systematic uncertainty. 
To assess whether such effects are supported by the data, we compare the parameters obtained with model ng1 when restricting the dataset to pre- and post-perihelion observations.

While the full-arc ng1 solution yields well-constrained NGA parameters, the split-arc solutions show substantially enlarged uncertainties and reduced stability (cf. Table~\ref{tab:orbfits}). 
In particular, the values of $A_1$ are inconsistent with that of the full-arc solution; $A_2$ becomes poorly constrained; $A_3$ is inconsistent between the pre- and post-perihelion solutions and with the full-arc solution. 
Overall, the degraded precision and the emergence of inter-solution discrepancies indicate that the split-arc configurations are significantly more sensitive to mutual parameter correlations and residual observational biases than the full-arc fit (see also the discussion in Section~\ref{deweight_sensitivity} below). 
Another possibility is that the discrepancies in the NGA parameters derived from the split-arc solutions reflect a degree of asymmetry not captured by the standard symmetric formulation \citep{Krolikowska_Dybczynski:2013}, or, more generally, the limitations inherent in assuming constant NGA parameters over the full observational arc.

Motivated by the results of the split-arc orbit fits, we fitted the data with the ng2 model, which introduces the delay parameter $DT$ as an additional degree of freedom. 
This model yields a statistically significant delay of several days in the peak outgassing, providing independent support for the NGA asymmetry suggested by the split-arc solutions.
As in the full-dataset ng1 case, however, the relatively small formal uncertainties do not exclude the presence of systematic biases affecting the inferred NGA parameters.

In all three panels of Figure~\ref{fig:residuals}, the distribution of residuals for the validation sample follows a qualitatively similar pattern to that of the fitted sample and, quantitatively, its RMS is only moderately larger. 
These features indicate that the orbital solutions derived from the fitted dataset are satisfactorily robust, and also provide a reasonable match to the validation data.

Among the Tier 2 observations included in the fit, approximately 31\% are rejected as outliers in the gr0 solution; this fraction decreases to about 20\% for both the ng1 and ng2 solutions. 
This substantial reduction in the rejection fraction relative to the gravity-only model indicates that the non-gravitational models provide a significantly better representation of the observations. 
It also suggests that using observer-provided uncertainties (i.e., without inflation) for Tier 2 observations is appropriate in this context. 
Tier 3 observations, on the other hand, are almost never rejected ($< 0.1\%$ in all cases), but their contribution to the fit is already reduced through the inflation of their reported uncertainties.

In summary, including symmetric NGA significantly improves the fit relative to a purely gravitational model. 
However, the discrepancy between pre- and post-perihelion solutions, and the improvement of the fit obtained with the delayed peak NGA model, indicate that a symmetric formulation may be incomplete. 
The inferred NGA parameters may therefore be affected by systematic variations arising from asymmetric outgassing. 
The dynamical consequences of such asymmetry for \trei{} are examined further in Section~\ref{asymm}.

\subsection{Sensitivity of the NGA parameters to the uncertainty inflation factor}
\label{deweight_sensitivity}

\begin{figure}
\begin{center}
\includegraphics[width=0.49\textwidth]{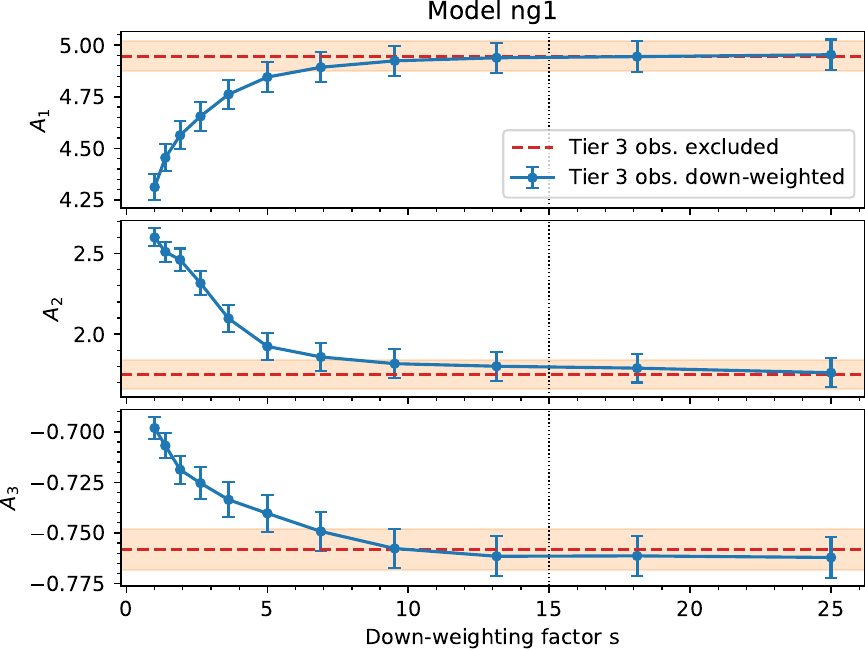}
\caption{Dependence of the NGA parameters on the scale factor $s$ for the full-arc ng1 solution. Blue circles with error bars show the fitted values obtained for each value of $s$. The red dashed line, with the shaded region indicating its uncertainty, corresponds to an orbital solution obtained by excluding Tier 3 observations. The black dotted vertical line marks the value $s=15$ adopted in our orbit fits.}
\label{fig:varys}
\end{center}
\end{figure}

We tested the robustness of the fitted NGA parameters with respect to the choice of the scale factor $s$ by repeating the full-arc ng1 fit for values of $s$ in the range $1 \le s \le 25$. 
The results are shown in Figure~\ref{fig:varys}. 
All three NGA parameters converge toward nearly constant values as $s$ increases. 
These asymptotic values are consistent, within uncertainties, with the solution obtained when Tier 3 observations are excluded from the fit. 
This behavior matches the theoretical expectation. 
Figure~\ref{fig:varys} also shows that convergence is effectively achieved for $s \gtrsim10$. 
We therefore adopt $s=15$, for which the solution is already insensitive to further increases in the uncertainty inflation factor.

\begin{figure}
\begin{center}
\includegraphics[width=0.49\textwidth]{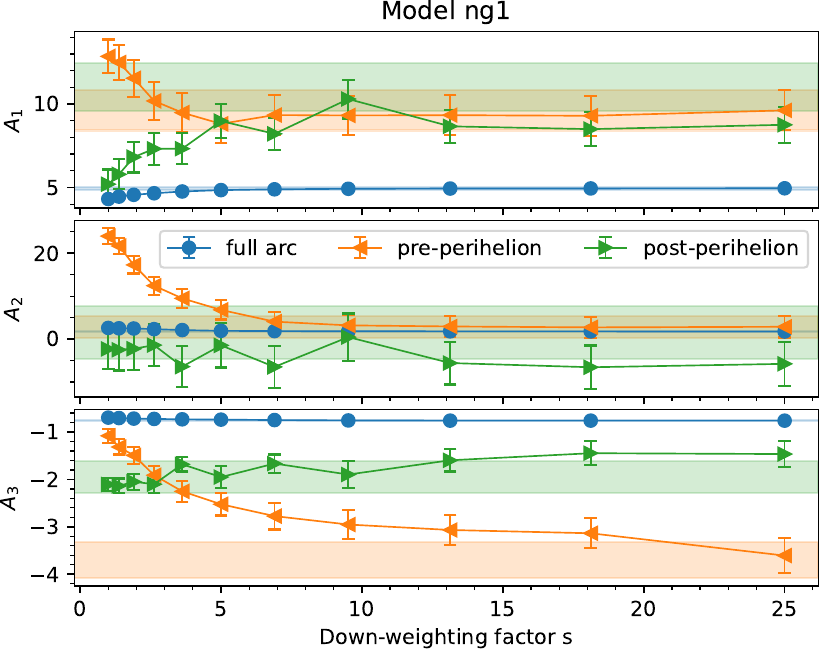}
\caption{
As in Figure~\ref{fig:varys}, but comparing the full-arc, pre-perihelion, and post-perihelion ng1 solutions (blue circles, left-pointing triangles, and right-pointing triangles, respectively). For each case, the shaded region of the corresponding color indicates the uncertainty range of the solution obtained by excluding Tier 3 observations.
}
\label{fig:varys_split}
\end{center}
\end{figure}

The NGA parameters determined in the pre- and post-perihelion ng1 fits were found in Section~\ref{orbfits} to be mutually inconsistent, as well as inconsistent with the full-arc solution.
To investigate the origin of these discrepancies, the dependence of the NGA parameters on the scale factor $s$ was also evaluated for the ng1-pre and ng1-pst fits (cf. Figure~\ref{fig:varys_split}). 

For both split-arc configurations, considered separately, we recover the asymptotic behavior with increasing $s$ observed in the full-arc ng1 fit, with the solutions approaching those obtained when Tier 3 observations are excluded. 
In the ng1-pst case, however, a small residual discrepancy (at the $\lesssim 1\sigma$ level) persists between the asymptotic NGA coefficients and those of the corresponding Tier~1+2-only solution.

On the other hand, the reduced constraint resulting from the absence of one arc leads to significantly enlarged uncertainties and reduced robustness of the NGA parameters.
Specifically, as already noted in Section~\ref{orbfits}, for $A_1$, both fits converge to mutually consistent values that are inconsistent with the full-arc solution; for $A_2$, the asymptotic values are broadly consistent with the full-arc result, but with substantially larger uncertainties, making $A_2$ compatible with zero; for $A_3$, the fits do not converge to a common asymptotic value and instead show divergent behavior.

The inconsistencies found in the split-arc solutions (mutual in the case of $A_3$, and with respect to the full-arc solution in the case of $A_1$) are at the $\gtrsim 4\sigma$ level. 
This tension suggests that residual tailward bias may still affect the solutions, particularly in the less stable split-arc fitting configurations. 
While a contribution from model limitations cannot be excluded, the relative importance of the different sources of systematic error cannot be robustly disentangled on the basis of the present analysis.

The results of the foregoing sensitivity analysis can be summarized as follows. Figure~\ref{fig:varys} illustrates that Tier 3 observations can contribute to systematic shifts in the derived NGA parameters if their uncertainties are not appropriately handled. 
Although the adopted uncertainty scheme substantially reduces the impact of astrometric bias, the behavior of the split-arc solutions in Figure~\ref{fig:varys_split} suggests that some residual systematic contributions may persist.
This highlights the inherent difficulty of fully mitigating systematic astrometric biases through uncertainty handling alone.

\subsection{Orbit fits with asymmetric NGA decay rate}
\label{asymm}

Observational evidence indicates that the cometary activity of \trei{} was significantly asymmetric  with respect to the perihelion in terms of the water production rate \citep{Tan_ea:2026}, as well as the activity index \citep{Hui_ea:2026}. 
Although other interpretations are possible, this observed behavior may point to an asymmetry in outgassing, and consequently in the radial decay of the NGA when moving away from the time of its peak.
To test this possibility, we fitted the \trei{} observations with model ng3, which implements an NGA scaling with the heliocentric distance according to equation \eqref{g_nga}.
In this dynamical model, the parameters $m$ and $\delta_m$ control the average exponent and the difference between the effective pre- and post-perihelion exponents, respectively.
To ensure internal consistency with the solutions presented in Section~\ref{orbfits}, the fits discussed below are referenced to the same epoch (\fitepoch{}) and include only observations up to the same cutoff date.

The parameters $m$ and $\delta_m$ of model ng3 are strongly correlated with the scale factor of the NGA parameters.
Specifically, larger values of $m$ require a larger overall NGA magnitude, $A$, to fit the same observations, due to the steeper radial decay of the acceleration with heliocentric distance.
This quasi-degeneracy can only be effectively broken if observational data sampling a wide range of heliocentric distances are available.
Preliminary tests showed that a least-squares fit encounters convergence issues, due to ill-conditioning of the normal matrix, if both $m$ and $\delta_m$ are varied at once.

\begin{table*}
\caption{Parameters of the ng3 orbital solutions discussed in Section~\ref{asymm} ($m_{\rm pre} = m - \delta_m$; $m_{\rm post} = m + \delta_m$).}
\begin{center}
\begin{tabular}{ccccccccc}
\hline
\hline
$m$ & res. RMS & $\chi^2_v$ & $m_{\rm pre}$ & $m_{\rm post}$ & $A_1$ & $A_2$ & $A_3$ & $A$ \\
        & (arc sec) & & & &  \multicolumn{4}{c}{($10^{-8}$ au d$^{-2}$)} \\
\hline
1.0 & $ 0.771$ & $ 0.059$ & $ 1.24 \pm 0.02$ & $ 0.76 \pm 0.02$ & $ 3.08 \pm 0.05$ & $ 1.39 \pm 0.09$ & $-0.42 \pm 0.01$ & $ 3.40 \pm 0.05$ \\
1.5 & $ 0.771$ & $ 0.058$ & $ 1.72 \pm 0.03$ & $ 1.28 \pm 0.03$ & $ 4.06 \pm 0.06$ & $ 1.30 \pm 0.11$ & $-0.57 \pm 0.01$ & $ 4.30 \pm 0.07$ \\
2.0 & $ 0.770$ & $ 0.059$ & $ 2.25 \pm 0.03$ & $ 1.75 \pm 0.03$ & $ 5.27 \pm 0.07$ & $ 1.18 \pm 0.14$ & $-0.75 \pm 0.01$ & $ 5.45 \pm 0.08$ \\
2.5 & $ 0.770$ & $ 0.059$ & $ 2.78 \pm 0.04$ & $ 2.22 \pm 0.04$ & $ 6.69 \pm 0.09$ & $ 1.09 \pm 0.18$ & $-0.97 \pm 0.01$ & $ 6.85 \pm 0.09$ \\
3.0 & $ 0.770$ & $ 0.059$ & $ 3.35 \pm 0.05$ & $ 2.65 \pm 0.05$ & $ 8.39 \pm 0.11$ & $ 0.94 \pm 0.23$ & $-1.24 \pm 0.02$ & $ 8.53 \pm 0.11$ \\
\hline
\end{tabular}
\end{center}
\label{tab:asymm}
\end{table*}

For these reasons, we do not attempt to estimate all parameters of model ng3 simultaneously, but rather sample a range of values for the average exponent $m$, fitting the remaining parameters for each choice. 
Our results are summarized in Table~\ref{tab:asymm}.
As expected, varying $m$ leads to systematic changes in the best-fitting values of $\delta_m$ and of the NGA coefficients. 
The slope difference parameter $\delta_m$ is negative in all cases, indicating a shallower radial decay of the NGA in the post-perihelion arc relative to the pre-perihelion arc.
The radial and normal components of the NGA, $A_1$ and $A_3$, increase monotonically with $m$ (the latter in absolute value).
As noted above, such a trend is expected: as $m$ increases, the function $g(r)$ becomes more sharply peaked, and the corresponding NGA amplitudes must increase to produce a comparable integrated effect over the observed trajectory.
In contrast, the transverse component, $A_2$, decreases with increasing $m$, contrary to this expectation.
 Its formal uncertainty is also larger than that of the other two components. 
 This behavior is consistent with $A_2$ being the least robustly constrained of the NGA parameters, likely reflecting a greater sensitivity to parameter degeneracies and systematic biases.

\begin{figure}
\begin{center}
\includegraphics[width=0.5\textwidth]{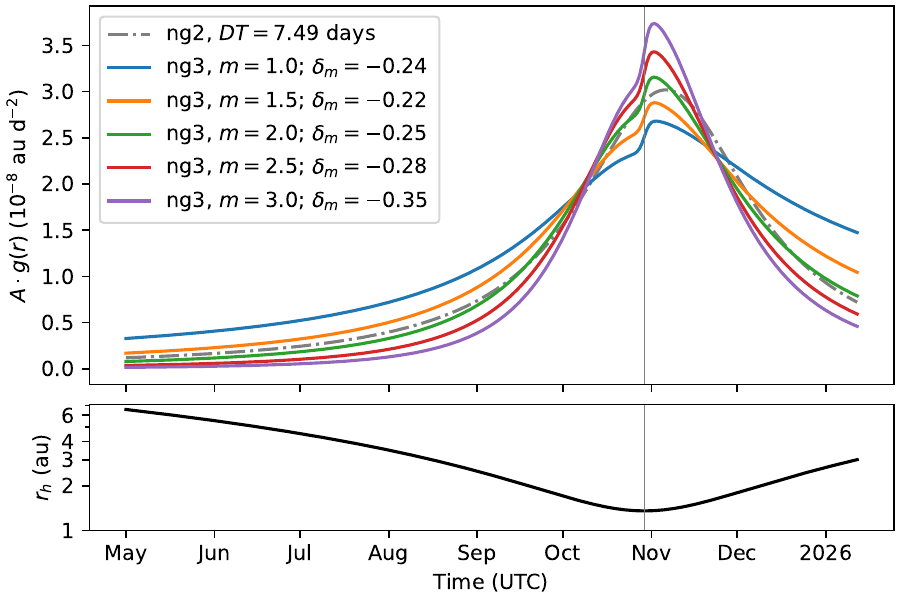}
\caption{Magnitude of the NGA as a function of time for the fits reported in Table~\ref{tab:asymm}. The bottom panel shows the corresponding variation of the heliocentric distance.}
\label{fig:nga}
\end{center}
\end{figure}

In spite of the significant variations in the fitted NGA parameters, the quality of the fits is almost insensitive to $m$.
The goodness-of-fit metrics are also very close to those of the ng1 and ng2 models discussed in the previous section.
The available data are therefore not very sensitive to the details of the radial dependence of the NGA.
As stated above, this behavior can be understood by noting that the observations span a relatively shallow range of heliocentric distances (approximately a factor of $4$). 
Within this interval, different combinations of the parameters can produce nearly identical integrated non-gravitational effects along the trajectory, as illustrated in Figure~\ref{fig:nga}.

In Figure~\ref{fig:nga}, the magnitude of the NGA for our ng3-type models is also compared with that obtained for the ng2 model discussed in Section~\ref{orbfits}. 
The Figure shows that the time evolution of $A \cdot g(r)$ for the ng3 model with $m=2$ is broadly similar to that of model ng2.
Consistently, the corresponding NGA parameters inferred from orbital solution ng2 and ng3 (with $m=2$) are very close. 
Moreover, all ng3 fits shown in the figure converge to values of  $m$ and $\delta_m$ that yield an NGA peak offset comparable to that of the ng2 solution, i.e., a delay $DT \approx 7.5$ days after perihelion. 
This suggests that, over the observed arc, variations in slope asymmetry can mimic an effective shift in the peak of the NGA, and vice versa, resulting in a partial degeneracy between the two parameterizations.

The additional degree of freedom, and accompanying degeneracy, arising from $\delta_m$ (or $DT$) introduces a further systematic source of uncertainty on the NGA coefficients and on the total acceleration amplitude $A$.
For the range of $m$ shown in Table~\ref{tab:asymm}, $A$ scales approximately as $2.1 \times (1.6)^m$. 

It would be very valuable to have an independent estimate of $m$ from observations.
In this respect, we note that the activity index of \trei{} was reported as $3.8 \pm 0.3$ pre-perihelion \citep{Jewitt_Luu:2025} and $4.5 \pm 0.3$ post-perihelion \citep{Hui_ea:2026}.
This apparent steepening of the activity profile is not reflected in our fitted NGA. 
However, the relationship between photometric activity and NGA is not necessarily straightforward. 
The reported water-production rates, even when derived from the same observational dataset, remain controversial \citep[see][]{Combi_ea:2026, Tan_ea:2026}. 
In particular, although the production rates derived by \citet{Tan_ea:2026} differ substantially from ours, they likewise exhibit a shallower dependence on heliocentric distance after perihelion, with power-law exponents of $5.9 \pm 0.8$ inbound and $3.3 \pm 0.3$ outbound. 
More generally, the physical processes driving the activity may evolve substantially during the apparition.

Recent studies suggest that \trei{} may have transitioned from CO$_2$-dominated activity before perihelion to H$_2$O-dominated activity afterward \citep{Moreno_ea:2026}. 
Such behavior is not explicitly represented by standard empirical NGA prescriptions, which assume a fixed functional dependence of the acceleration on heliocentric distance. 
Consequently, even in the absence of astrometric biases, model-dependent systematic uncertainties are expected to remain. 
This provides a plausible explanation for part of the dispersion among acceptable NGA solutions and reinforces the need to account for modeling systematics when deriving physical properties from the inferred acceleration.

\section{Discussion}
\label{discussion}

Our estimates of the NGA coefficients can be directly compared with orbital solution \#54 from the JPL Small-Body Database (SBDB)\footnote{\url{https://ssd.jpl.nasa.gov/tools/sbdb_lookup.html\#/?sstr=3i}}, which corresponds to an ng2-type model with $g(r) = (1 {\rm au}/r)^2$. 
Its best-fitting NGA parameters are $A_1 = 5.320 \pm 0.122$ au d$^{-2}$, $A_2 = 1.148 \pm 0.214$ au d$^{-2}$, $A_3 = -0.6854 \pm 0.0178$ au d$^{-2}$, and $DT = 9.479 \pm 1.426$ days.

Comparing these values with those reported in Table~\ref{tab:orbfits} for our ng2 solution, we find that the two solutions are essentially consistent within their formal uncertainties, with the exception of $A_3$, for which a discrepancy is observed at the $\sim 3\sigma$ level.
For comparison, our (full-arc) ng1 solution shows larger deviations with respect to an earlier symmetric NGA solution from the JPL SBDB (\#31, no longer publicly accessible), most notably in $A_1$ ($\sim 3\sigma$) and $A_3$ ($\sim 6\sigma$).
We note that parameter correlations may cause different NGA components to partially absorb mismodeling or residual systematics affecting other components.

Given the similar treatment of the NGA, these discrepancies are plausibly attributable to differences in the preprocessing of the observations, in particular the treatment of tailward bias, as well as in the adopted error model. 
The JPL SBDB solution \#54 quoted above is based on $782$ observations, indicating a substantially vetted dataset relative to that publicly available from the MPC.

The smaller formal uncertainties of our solution compared to the JPL SBDB solution are consistent with the larger number of observations included in our analysis and may also reflect differences in the adopted uncertainty model. 
At the same time, the sensitivity tests presented in Section~\ref{deweight_sensitivity} indicate that some residual systematic effects associated with lower-quality astrometry may persist. 
Consequently, the quoted formal uncertainties of our solution should not be interpreted as fully capturing the uncertainty in the derived NGA parameters.

Additional determinations of the NGA parameters were presented by \citet{Cloete_ea:2025} and \citet{Ahuja_Ganesh:2025, Ahuja_Ganesh:2026}. 
However, these analyses relied on more limited datasets, notably lacking post-perihelion observations, and employed NGA models that differ significantly from ours. 
For these reasons, we do not attempt a direct comparison with those results.

\citet{Eubanks_ea:2025} reported orbital fits for \trei{} that include NGA effects. 
The solutions labeled “E+P+TGO (3 d.f.)” and “E+P+TGO (4 d.f.)” in their Table~1 correspond most closely to our models ng1 and ng2, respectively. 
In both cases, significant quantitative and qualitative differences are evident. 
In particular, their four-degree-of-freedom solution yields negative best-fitting values of both $A_1$ and $DT$\footnote{Note that their adopted definition of $DT$ has the opposite sign with respect to ours.}. 
A negative radial NGA component is difficult to reconcile with the conventional sublimation-driven interpretation of cometary activity. 
Their preferred DT value furthermore implies peak outgassing occurring approximately one month before perihelion, but their observational arc extends only about one month beyond perihelion. 
Again, these discrepancies are likely due to differences in the observational dataset and in the adopted uncertainty model.

In summary, among the previously published NGA determinations for \trei{}, the solution provided by JPL's SBDB is overall the most consistent with ours.

In this work, we identified and quantified the main sources of uncertainty affecting the determination of the NGA parameters for \trei{}. 
Our preferred orbital solutions are obtained using the two most common approaches to modeling the NGA: a radial dependence symmetric with respect to perihelion (model ng1), and an asymmetric dependence implemented as a freely adjustable time offset in the peak outgassing (model ng2).
We also obtain a comparably satisfactory fit with an alternative prescription that assumes an asymmetric radial decay of the outgassing rate away from its maximum (model ng3).

Accurate determination of the NGA parameters is essential for estimating the nucleus size of \trei{}. 
All other quantities being equal, the radius scales with the magnitude of the NGA normalized at $1$ au approximately as $\frac{\delta R}{R} \sim \frac{1}{3} \frac{\delta A}{A}$ (see equation 10 of \citealt{Hui_ea:2020}).
Our estimate of $A$ for our ng2 solution is consistent with that of the JPL SBDB solution \#54 ($5.493 \pm 0.084$ vs $5.55 \pm 0.087$, both in units of $10^{-8}$ au d$^{-2}$).
If the additional systematic effects on $A$ associated with our asymmetric radial decay models ng3 are included (see Section~\ref{asymm}), however, the corresponding relative uncertainty in the radius is $\approx 15\%$.

\section{Conclusions}
\label{conclusions}

We presented an analysis of the trajectory of interstellar comet \trei{}, with particular emphasis on the determination of its non-gravitational acceleration (NGA) and the assessment of the associated uncertainties.

Our preferred orbital solution (ng2), featuring an asymmetry in the NGA relative to perihelion parametrized by a delayed peak in the outgassing law, is directly comparable to the JPL solution \#54. 
The two solutions are broadly consistent, with the largest discrepancy occurring in the normal component, $A_3$.
This agreement, together with the stability of the inferred total acceleration across the different models considered here, suggests that the inferred NGA of \trei{} is sufficiently robust to support physical interpretation and nucleus-size estimates.

The comparison of orbital fits obtained from different subsets of the astrometric data highlights the impact of residual tailward biases on NGA determinations. 
Symmetric NGA models fitted separately to the pre-perihelion and post-perihelion arcs yield inconsistent NGA coefficients and differ significantly from the full-arc solution. 
This indicates that NGA estimates derived from short observational arcs may be strongly biased unless astrometric systematics are properly mitigated. 
The uncertainty assignment scheme adopted in this work reduces, but does not fully remove, the influence of such systematics on the pre- and post-perihelion solutions.

At the same time, our analysis underscores the limitations of simplified empirical descriptions of cometary activity. 
Alternative parameterizations of the outgassing law, including asymmetric activity slopes relative to perihelion, yield fits of comparable quality to the standard ng2 model but lead to systematic differences in the inferred NGA parameters.
These model-dependent effects contribute significantly to the uncertainty budget and exceed the formal statistical errors.
As a result, covariance-based uncertainties likely underestimate the true errors on derived physical quantities, including the nucleus size of \trei{}.

\begin{acknowledgements}
The authors gratefully acknowledge the anonymous referee, whose careful review and constructive suggestions greatly improved the manuscript.
This research has made use of positional data of \trei{} provided by the International Astronomical Union’s Minor Planet Center.
\end{acknowledgements}

\bibliographystyle{aa}

\begin{thebibliography}{}

\bibitem[Acton(1996)]{Acton_ea:1996} Acton, C.~H.\ 1996, \planss, 44, 1, 65. doi:10.1016/0032-0633(95)00107-7

\bibitem[Acton et al.(2018)]{Acton_ea:2018} Acton, C., Bachman, N., Semenov, B., et al.\ 2018, \planss, 150, 9. doi:10.1016/j.pss.2017.02.013

\bibitem[Ahuja \& Ganesh(2025)]{Ahuja_Ganesh:2025} Ahuja, G. \& Ganesh, S.\ 2025, \apjl, 995, 1, L13. doi:10.3847/2041-8213/ae21cf

\bibitem[Ahuja \& Ganesh(2026)]{Ahuja_Ganesh:2026} Ahuja, G. \& Ganesh, S.\ 2026, Research Notes of the American Astronomical Society, 10, 1, 19. doi:10.3847/2515-5172/ae3c09

\bibitem[Albrow et al.(2025)]{Albrow_ea:2025} Albrow, L., Bannister, M.~T., Forbes, J.~C., et al.\ 2025, arXiv:2512.04700. doi:10.48550/arXiv.2512.04700

\bibitem[Annex et al.(2020)]{Annex_ea:2020} Annex, A., Pearson, B., Seignovert, B., et al.\ 2020, The Journal of Open Source Software, 5, 46, 2050. doi:10.21105/joss.02050

\bibitem[Bolin et al.(2025)]{Bolin_ea:2025} Bolin, B.~T., Belyakov, M., Fremling, C., et al.\ 2025, \mnras, 542, 1, L139. doi:10.1093/mnrasl/slaf078

\bibitem[Carpino et al.(2003)]{Carpino_ea:2003} Carpino, M., Milani, A., \& Chesley, S.~R.\ 2003, \icarus, 166, 2, 248. doi:10.1016/S0019-1035(03)00051-4

\bibitem[Cloete et al.(2025)]{Cloete_ea:2025} Cloete, R., Loeb, A., \& Vere{\v{s}}, P.\ 2025, arXiv:2509.21408. doi:10.48550/arXiv.2509.21408

\bibitem[Combi et al.(2026)]{Combi_ea:2026} Combi, M.~R., M{\"a}kinen, T., Bertaux, J.~L., et al.\ 2026, \apjl, 998, 1, L17. doi:10.3847/2041-8213/ae3bd8

\bibitem[Cordiner et al.(2025)]{Cordiner_ea:2025} Cordiner, M.~A., Roth, N.~X., Kelley, M.~S.~P., et al.\ 2025, \apjl, 991, 2, L43. doi:10.3847/2041-8213/ae0647

\bibitem[Cordiner et al.(2026)]{Cordiner_ea:2026} Cordiner, M., Roth, N.~X., Micheli, M., et al.\ 2026, \nat, 655, 8124, 870. doi:10.1038/s41586-026-10771-6

\bibitem[Damour \& Deruelle(1985)]{Damour_Deruelle:1985} Damour, T. \& Deruelle, N.\ 1985, Annales de L'Institut Henri Poincare Section (A) Physique Theorique, 43, 1, 107. 

\bibitem[Denneau et al.(2025)]{Denneau_ea:2025} Denneau, L., Siverd, R., Tonry, J., et al.\ 2025, Minor Planet Electronic Circulars, 2025-N12. doi:10.48377/MPEC/2025-N12

\bibitem[Eggl et al.(2020)]{Eggl_ea:2020} Eggl, S., Farnocchia, D., Chamberlin, A.~B., et al.\ 2020, \icarus, 339, 113596

\bibitem[Eubanks et al.(2025)]{Eubanks_ea:2025} Eubanks, T.~M., Hibberd, A., Bills, B.~G., et al.\ 2025, Research Notes of the American Astronomical Society, 9, 12, 329. doi:10.3847/2515-5172/ae2915

\bibitem[Farnocchia et al.(2015)]{Farnocchia_ea:2015} Farnocchia, D., Chesley, S.~R., Milani, A., et al.\ 2015, Asteroids IV, 815. doi:10.2458/azu\_uapress\_9780816532131-ch041

\bibitem[Farnocchia et al.(2016)]{Farnocchia_ea:2016} Farnocchia, D., Chesley, S.~R., Micheli, M., et al.\ 2016, \icarus, 266, 279. doi:10.1016/j.icarus.2015.10.035

\bibitem[Fitzsimmons et al.(2018)]{Fitzsimmons_ea:2018} Fitzsimmons, A., Snodgrass, C., Rozitis, B., et al.\ 2018, Nature Astronomy, 2, 133. doi:10.1038/s41550-017-0361-4

\bibitem[Forbes \& Butler(2026)]{Forbes_Butler:2026} Forbes, J.~C. \& Butler, H.\ 2026, Research Notes of the American Astronomical Society, 10, 1, 12. doi:10.3847/2515-5172/ae3747

\bibitem[Guo et al.(2025)]{Guo_ea:2025} Guo, Y., Zhang, L., Feng, F., et al.\ 2025, \aj, 170, 6, 362. doi:10.3847/1538-3881/ae1833

\bibitem[Holman et al.(2023)]{Holman_ea:2023} Holman, M.~J., Akmal, A., Farnocchia, D., et al.\ 2023, \psj, 4, 4, 69. doi:10.3847/PSJ/acc9a9

\bibitem[Hopkins et al.(2025)]{Hopkins_ea:2025} Hopkins, M.~J., Dorsey, R.~C., Forbes, J.~C., et al.\ 2025, \apjl, 990, 2, L30. doi:10.3847/2041-8213/adfbf4

\bibitem[Hui et al.(2020)]{Hui_ea:2020} Hui, M.-T., Ye, Q.-Z., F{\"o}hring, D., et al.\ 2020, \aj, 160, 2, 92. doi:10.3847/1538-3881/ab9df8

\bibitem[Hui et al.(2026)]{Hui_ea:2026} Hui, M.-T., Jewitt, D., Mutchler, M.~J., et al.\ 2026, \apjl, 999, 2, L37. doi:10.3847/2041-8213/ae471c

\bibitem[Jewitt et al.(2017)]{Jewitt_ea:2017} Jewitt, D., Luu, J., Rajagopal, J., et al.\ 2017, \apjl, 850, 2, L36. doi:10.3847/2041-8213/aa9b2f

\bibitem[Jewitt(2024)]{Jewitt:2024} Jewitt, D. \ (2024). Interstellar Objects in the Solar System. In: Deeg, H.J., Belmonte, J.A. (eds) Handbook of Exoplanets . Springer, Cham, Switzerland. doi:10.1007/978-3-319-30648-3\_197-1

\bibitem[Jewitt et al.(2025)]{Jewitt_ea:2025} Jewitt, D., Hui, M.-T., Mutchler, M., et al.\ 2025, \apjl, 990, 1, L2. doi:10.3847/2041-8213/adf8d8

\bibitem[Jewitt \& Luu(2025)]{Jewitt_Luu:2025} Jewitt, D. \& Luu, J.\ 2025, \apjl, 994, 1, L3. doi:10.3847/2041-8213/ae1832

\bibitem[Jewitt \& Seligman(2023)]{Jewitt_Seligman:2023} Jewitt, D. \& Seligman, D.~Z.\ 2023, \araa, 61, 197. doi:10.1146/annurev-astro-071221-054221

\bibitem[Kim et al.(2020)]{Kim_ea:2020} Kim, Y., Jewitt, D., Mutchler, M., et al.\ 2020, \apjl, 895, 2, L34. doi:10.3847/2041-8213/ab9228

\bibitem[Kr{\'o}likowska \& Dones(2023)]{Krolikowska_Dones:2023} Kr{\'o}likowska, M. \& Dones, L.\ 2023, \aap, 678, A113. doi:10.1051/0004-6361/202347178

\bibitem[Kr{\'o}likowska \& Dybczy{\'n}ski(2013)]{Krolikowska_Dybczynski:2013} Kr{\'o}likowska, M. \& Dybczy{\'n}ski, P.~A.\ 2013, \mnras, 435, 1, 440. doi:10.1093/mnras/stt1313

\bibitem[Martinez-Palomera et al.(2025)]{Martinez-Palomera_ea:2025} Martinez-Palomera, J., Tuson, A., Hedges, C., et al.\ 2025, \apjl, 994, 2, L51. doi:10.3847/2041-8213/ae1f91

\bibitem[Marsden et al.(1973)]{Marsden_ea:1973} Marsden, B.~G., Sekanina, Z., \& Yeomans, D.~K.\ 1973, \aj, 78, 211. doi:10.1086/111402

\bibitem[Meech et al.(2017)]{Meech_ea:2017} Meech, K.~J., Weryk, R., Micheli, M., et al.\ 2017, \nat, 552, 7685, 378. doi:10.1038/nature25020

\bibitem[Micheli et al.(2018)]{Micheli_ea:2018} Micheli, M., Farnocchia, D., Meech, K.~J., et al.\ 2018, \nat, 559, 223. doi:10.1038/s41586-018-0254-4

\bibitem[Milani \& Gronchi(2010)]{Milani_Gronchi:2010} Milani, A. and Gronchi, G., 2010. \textit{Theory of Orbit Determination} Cambridge University Press.

\bibitem[Moreno et al.(2026)]{Moreno_ea:2026} Moreno, F., Serra-Ricart, M., Licandro, J., et al.\ 2026, \mnras, 550, 1, stag1164. doi:10.1093/mnras/stag1164

\bibitem[Moro-Mart{\'\i}n(2022)]{Moro-Martin:2022} Moro-Mart{\'i}n, A. \ 2023: In  Planetary Systems Now, ed. by Luisa Lara and David Jewitt, World Scientific, 2023

\bibitem[Park et al.(2021)]{Park_ea:2021} Park, R.~S., Folkner, W.~M., Williams, J.~G., et al.\ 2021, \aj, 161, 3, 105. doi:10.3847/1538-3881/abd414

\bibitem[Rein \& Liu(2012)]{Rein_Liu:2012} Rein, H. \& Liu, S.-F.\ 2012, \aap, 537, A128. doi:10.1051/0004-6361/201118085

\bibitem[Rein \& Spiegel(2015)]{Rein_Spiegel:2015} Rein, H. \& Spiegel, D.~S.\ 2015, \mnras, 446, 2, 1424. doi:10.1093/mnras/stu2164

\bibitem[Salazar Manzano et al.(2026)]{SalazarManzano_ea:2026} Salazar Manzano, L.~E., Paneque-Carre{\~n}o, T., Cordiner, M.~A., et al.\ 2026, Nature Astronomy. doi:10.1038/s41550-026-02850-5

\bibitem[Seligman et al.(2025)]{Seligman_ea:2025} Seligman, D.~Z., Micheli, M., Farnocchia, D., et al.\ 2025, \apjl, 989, 2, L36. doi:10.3847/2041-8213/adf49a

\bibitem[Shampine \& Gordon(1975)]{Shampine_Gordon:1975} Shampine, L.~F., \& Gordon, M.~K. 1975, \textit{Computer Solution of Ordinary Differential Equations: The Initial Value Problem} (San Francisco, CA: W.~H. Freeman)

\bibitem[Spada(2024)]{orbitfinder} Spada, F. 2024, \texttt{orbit\_finder}, GitHub repository, \url{https://github.com/federico-spada/orbit_finder}

\bibitem[Tan et al.(2026)]{Tan_ea:2026} Tan, H., Yan, X., \& Li, J.-Y.\ 2026, \apjl, 998, 1, L22. doi:10.3847/2041-8213/ae3c97

\bibitem[Taylor \& Seligman(2025)]{Taylor_Seligman:2025} Taylor, A.~G. \& Seligman, D.~Z.\ 2025, \apjl, 990, 1, L14. doi:10.3847/2041-8213/adfa28

\bibitem[Tholen \& Chesley(2004)]{Tholen_Chesley:2004} Tholen, D.~J. \& Chesley, S.~R.\ 2004, American Astronomical Society, DPS meeting \#36, 36, 34.16. 

\bibitem[Vere{\v{s}} et al.(2017)]{Veres_ea:2017} Vere{\v{s}}, P., Farnocchia, D., Chesley, S.~R., et al.\ 2017, \icarus, 296, 139. 

\bibitem[Vavilov et al.(2026)]{Vavilov_ea:2026} Vavilov, D.~E., Liu, Z., Hestroffer, D., et al.\ 2026, \icarus, 454, 117074. doi:10.1016/j.icarus.2026.117074

\bibitem[Watts(1983)]{Watts:1983} Watts, H.~A. 1983, Journal of Computational and Applied Mathematics, 9, 177, doi:10.1016/0377-0427(83)90040-7

\bibitem[Watts \& Shampine(1986)]{Watts_Shampine:1986} Watts, H.~A., \& Shampine, L.~F. 1986, SIAM Journal on Scientific and Statistical Computing, 7, 334, doi:10.1137/0907022

\bibitem[Yang et al.(2026)]{Yang_ea:2026} Yang, W., Yan, J., Chen, M., et al.\ 2026, \aap, 710, A8. doi:10.1051/0004-6361/202659299

\bibitem[Ye et al.(2017)]{Ye_ea:2017} Ye, Q.-Z., Zhang, Q., Kelley, M.~S.~P., et al.\ 2017, \apjl, 851, 1, L5. doi:10.3847/2041-8213/aa9a34

\bibitem[Yeomans \& Chodas(1989)]{Yeomans_Chodas:1989} Yeomans, D.~K. \& Chodas, P.~W.\ 1989, \aj, 98, 1083. doi:10.1086/115198

\end{thebibliography}

\begin{appendix}

\section{Osculating elements of our orbital fits}
\label{elts}

Table~\ref{tab:elts} reports the osculating heliocentric elements for the orbital solution gr0, ng1, and ng2 (see Section~\ref{orbfits}).
All the orbital elements in the table are referred to the ecliptic plane.
The tabulated values are intended to provide a reproducible numerical representation of the orbital solutions, enabling direct comparison among the different fits presented in this work, as well as comparison with independent determinations available in the literature.

Orbital elements correspond to osculating quantities at the stated fit epoch. 
In particular, the mean anomaly is defined consistently with the adopted dynamical model and reference frame. 
Care should therefore be exercised when comparing these values with solutions obtained under different modeling assumptions (e.g., planetary ephemerides, relativistic corrections, or non-gravitational force parameterizations), as such differences may result in small but non-negligible offsets even at the same epoch.

\begin{table*}
\centering
\caption{Heliocentric osculating orbital elements for the gr0, ng1, and ng2 orbital solutions.}
\begin{tabular}{l c c c}
\hline
\hline
Element & gr0 & ng1 & ng2 \\
\hline
$e$ & $6.1415349 \pm 2.20 \times 10^{-6}$ & $6.141359 \pm 1.10 \times 10^{-5}$ & $6.141330 \pm 1.50 \times 10^{-5}$ \\
$a$~(au) & $-0.263842560 \pm 8.50 \times 10^{-8}$ & $-0.26383911 \pm 3.10 \times 10^{-7}$ & $-0.26383836 \pm 3.90 \times 10^{-7}$ \\
$q$~(au) & $1.35655574 \pm 1.50 \times 10^{-7}$ & $1.3564917 \pm 1.50 \times 10^{-6}$ & $1.3564801 \pm 2.40 \times 10^{-6}$ \\
$i$~(deg) & $175.1157224 \pm 4.40 \times 10^{-6}$ & $175.116496 \pm 2.00 \times 10^{-5}$ & $175.116525 \pm 2.70 \times 10^{-5}$ \\
$\Omega$~(deg) & $322.167939 \pm 3.30 \times 10^{-5}$ & $322.169712 \pm 4.40 \times 10^{-5}$ & $322.169807 \pm 7.40 \times 10^{-5}$ \\
$\omega$~(deg) & $128.022918 \pm 3.40 \times 10^{-5}$ & $128.023195 \pm 4.80 \times 10^{-5}$ & $128.023038 \pm 6.10 \times 10^{-5}$ \\
$M$~(deg) & $818.19866 \pm 3.90 \times 10^{-4}$ & $818.2134 \pm 1.30 \times 10^{-3}$ & $818.2165 \pm 1.60 \times 10^{-3}$ \\
$A_1$~($10^{-8}$ au d$^{-2}$) & --- & $4.941 \pm 0.073$ & $5.381 \pm 0.085$ \\
$A_2$~($10^{-8}$ au d$^{-2}$) & --- & $1.801 \pm 0.089$ & $1.15 \pm 0.14$ \\
$A_3$~($10^{-8}$ au d$^{-2}$) & --- & $-0.762 \pm 1.00 \times 10^{-2}$ & $-0.7597 \pm 9.80 \times 10^{-3}$ \\
$DT$~(days) & --- & --- & $7.49 \pm 0.92$ \\
\hline
\end{tabular}
\label{tab:elts}
\end{table*}

\section{Extended residual diagnostics}
\label{addfig}

\begin{figure}
\begin{center}
\includegraphics[width=0.49\textwidth]{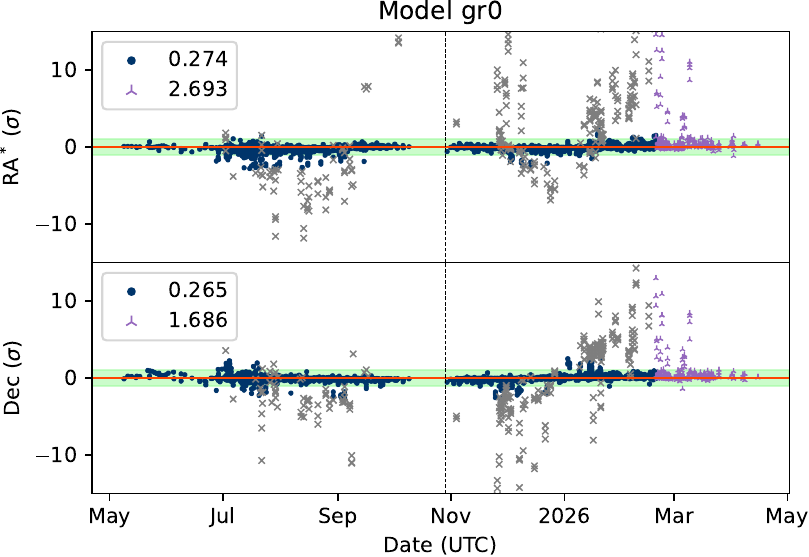}
\caption{Same as top panel of Figure~\ref{fig:residuals}, but with enlarged vertical axis limits.}
\label{fig:residuals_wide1}
\end{center}
\end{figure}

\begin{figure}
\begin{center}
\includegraphics[width=0.49\textwidth]{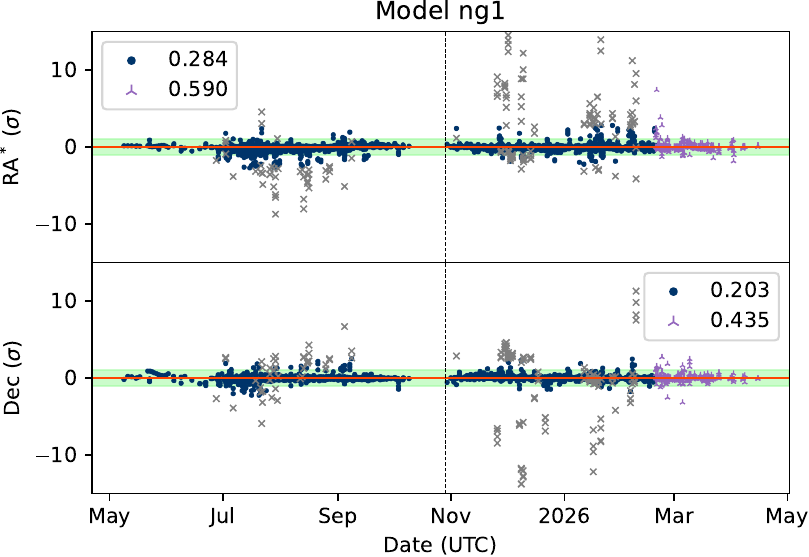}
\caption{Same as middle panel of Figure~\ref{fig:residuals}, but with enlarged vertical axis limits.}
\label{fig:residuals_wide2}
\end{center}
\end{figure}

\begin{figure}
\begin{center}
\includegraphics[width=0.49\textwidth]{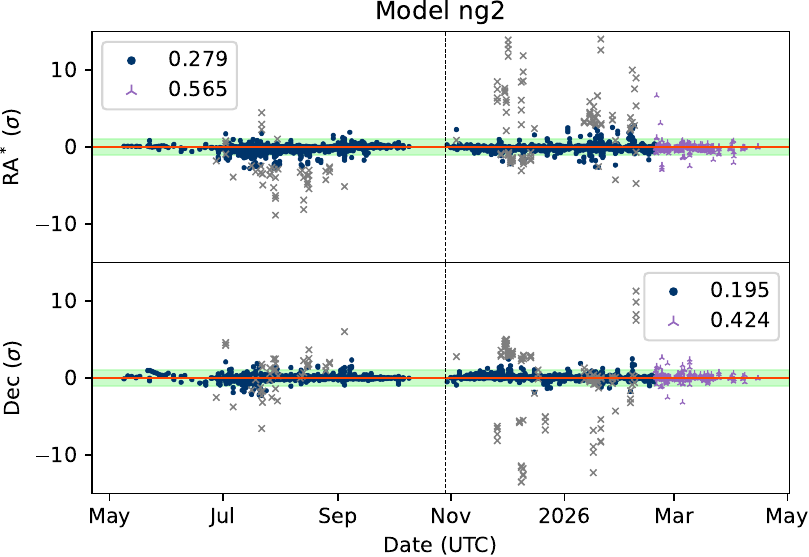}
\caption{Same as bottom panel of Figure~\ref{fig:residuals}, but with enlarged vertical axis limits.}
\label{fig:residuals_wide3}
\end{center}
\end{figure}

Figures~\ref{fig:residuals_wide1}, \ref{fig:residuals_wide2}, and \ref{fig:residuals_wide3} show expanded versions of the panels of Figure~\ref{fig:residuals}, identical in all respects except for wider vertical axis limits. 
This version allows visualization of the most extreme residuals, which fall outside the range of the plots shown in the main text.

\end{appendix}

\end{document}